# Spin polarizabilities and characteristics of spin-one hadrons related parity non-conservation in the the Duffin-Kemmer-Petiau formalism.

## N.V. Maksimenko[1], E.V. Vakulina[2],

[1]The F. Skorina Gomel State University, Gomel, Belarus

[2]Affiliated branch of Bryansk State University n.a. Academy Fellow I.G. Petrovskiy, Novozybkov, Russia

[1]**Abstract**. In this paper the method of the relativistically-invariant effective tensor representation of Lagrangian decomposition and the amplitudes of the two-photon interaction with hadrons on photon energy is proposed. In the frames Duffin-Kemmer-Petiau formalism based on this method the spin polarizabilities of the spin 1 particles, which are characteristic for the spin 1/2 hadrons were established. Also the authors have obtained the new spin polarizabilities which are characteristic for spin 1 particles and are bounded with the presence of the tensor polarizabilities of these particles. It was shown that in the suggested covariant approach with regard to the crossing symmetry, the spatial parity, conservation laws and gauge invariance definite spin polarizabilities and gyrations of spin-1 particles contribute to the expansion of Compton scattering amplitude, starting from the corresponding orders on photon energy that is in the agreement with low-energy theorems for that process.

## Introduction

On the basis of the low-energy theorems [1, 2] electromagnetic characteristics of nucleon - spin polarizabilities were established, which reflect the structural properties of the electromagnetic field interaction with these hadrons at low energies. The polarizabilities of hadrons have recently received much attention [1-10]. With the purpose of their experimental and theoretical study a fairly broad class of electrodynamic two-photon processes is used.

It was found that along with the dipole polarizabilities the interaction of electromagnetic field with hadrons can be determined also by the spin polarizabilities [1, 2].

With the development of of the standard model of the electroweak interactions recently new electromagnetic characteristics associated with non-conservation of parity [11, 12] similar to gyration [13] have been introduced.

To perfect the methods of improving the accuracy of measurement of polarizabilities, gyration and fitting the experimental data on the Compton scattering on hadrons at low and medium energies there is a need of a relativistic field-theoretical determination of the contributions of these characteristics in amplitude and cross section of electrodynamic processes [14-16].

For this purpose in this paper we propose a method of effective relativistic invariant tensor representation of Lagrangian decomposition and the amplitudes of

---

the two-photon interaction with hadrons on the photon energy [17-21]. As part of the proposed method and using the Duffin-Kemmer-Petiau formalism the new spin polarizabilities of of the spin 1 particles, which are determined by the tensor polarizability.

Also were obtained the characteristics of the spin 1 particles bounded with parity violation.

### Effective Lagrangian and Compton scattering amplitude in the formalism of the Duffin-Kemmer-Petiau considering spin polarizabilities.

We define relativistic invariant spin structures of the effective Lagrangian and Compton scattering amplitude on a particle of spin one in the formalism of the Duffin-Kemmer-Petiau (DKP) following [20-21].

DKP equation for a free particle of spin one has the form [22]:

$$(\beta_\mu \vec{\partial}_\mu + m)\psi(x) = 0, \tag{1}$$
$$\bar{\psi}(x)(\beta_\mu \overleftarrow{\partial}_\mu - m) = 0, \tag{2}$$

where $\psi(x)$ and $\bar{\psi}(x) = \psi^+(x)\eta$ - are ten-particle functions, $\eta = 2\left(\beta_4^{(10)}\right)^2 - I$, the arrows over derivatives $\partial_\mu$ indicate the direction of their actions and four-dimensional vector is determined by the components $a_\mu\{\vec{a}, ia_0\}$. In the equations (1) and (2) $\beta_\mu$ are ten-dimensional matrices of DKP, which satisfy the commutation relations:

$$\beta_\mu \beta_\nu \beta_\rho + \beta_\rho \beta_\nu \beta_\mu = \delta_{\mu\nu}\beta_\rho + \delta_{\rho\nu}\beta_\mu.$$

Effective Lagrangian of the interaction of electromagnetic fields with spin one particle considering polarizabilities within field-theoretical covariant approach has the form [17, 20]:

$$L = -\frac{\pi}{2m}\bar{\psi}[\beta_\nu \hat{L}_{\nu\sigma}\overleftrightarrow{\partial}_\sigma + \hat{L}_{\nu\sigma}\beta_\nu \overleftrightarrow{\partial}_\sigma]\psi, \tag{3}$$

where $\overleftrightarrow{\partial}_\sigma = \vec{\partial}_\sigma - \overleftarrow{\partial}_\sigma$.

In the definition of the Lagrangian (3) $\hat{L}_{\nu\sigma}$ tensor is expressed through polarizabilities and gyrations as follows:

$$\hat{L}_{\nu\sigma}(\alpha, \chi_E, \gamma_E) = \hat{L}_{\nu\sigma}(\alpha) + \hat{L}_{\nu\sigma}(\bar{\bar{\alpha}}) + \hat{L}_{\nu\sigma}(\gamma_{E_1}) + \hat{L}_{\nu\sigma}(\chi_E) + \hat{L}_{\nu\sigma}(\gamma_{E_2}), \tag{4}$$

$$\hat{L}_{\nu\sigma}(\beta, \chi_M, \gamma_M) = \hat{L}_{\nu\sigma}(\beta) + \hat{L}_{\nu\sigma}(\bar{\bar{\beta}}) + \hat{L}_{\nu\sigma}(\gamma_{M_1}) + \hat{L}_{\nu\sigma}(\chi_M) + \hat{L}_{\nu\sigma}(\gamma_{M_2}). \tag{5}$$

In the definitions (4) and (5) $\alpha$ and $\beta$ are the scalar dipole electric and magnetic polarizabilities, $\bar{\bar{\alpha}}$ and $\bar{\bar{\beta}}$ are the tensor polarizabilities, $\gamma_{E_1}, \gamma_{E_2}$ and

$\gamma_{M_1}, \gamma_{M_2}\ \gamma_{M_1}$ are the spin polarizabilities and $\mathcal{X}_E$ and $\mathcal{X}_M$ are the electrical and magnetic gyrations.

In order to establish the crossing symmetry influence on the contributions of the spin polarizabilities and gyrations in Compton scattering amplitude in the dipole representation we define tensors in (4) following [20]:

$$\hat{L}_{\nu\sigma}(\alpha) + \hat{L}_{\nu\sigma}(\bar{\bar{\alpha}}) = F_{\nu\mu}\hat{\alpha}_{\mu\rho}(\alpha)F_{\rho\sigma} + F_{\nu\mu}\hat{\alpha}_{\mu\rho}(\bar{\bar{\alpha}})F_{\rho\sigma}, \tag{6}$$

$$\hat{L}_{\nu\sigma}(\gamma_{E_1}) + \hat{L}_{\nu\sigma}(\chi_E) = F_{\nu\mu}\overleftrightarrow{\partial}_\lambda F_{\rho\sigma}\hat{k}_{\mu\rho\lambda}(\gamma_{E_1}) + F_{\nu\mu}\overleftrightarrow{\partial}_\lambda F_{\rho\sigma}\hat{k}_{\mu\rho\lambda}(\chi_E), \tag{7}$$

$$\hat{L}_{\nu\sigma}(\gamma_{E_2}) = (F_{\nu\rho}\overleftarrow{\partial}_\kappa \tilde{F}_{\sigma\rho} + \tilde{F}_{\nu\rho}\overrightarrow{\partial}_\rho F_{\sigma\kappa})\hat{k}_{\mu\rho\lambda}(\gamma_{E_2}). \tag{8}$$

The derivatives in (7-8) are effective only on the electromagnetic field tensors

$$F_{\mu\nu} = \partial_\mu A_\nu - \partial_\nu A_\mu.$$

Tensors $\hat{\alpha}^{\mu\rho}(\alpha)$ and $\hat{\alpha}^{\mu\rho}(\bar{\bar{\alpha}})$, as well as $\hat{k}_{\mu\rho\lambda}(\gamma_{E_1})$, $\hat{k}_{\mu\rho\lambda}(\gamma_{E_2})$ and $\hat{k}_{\mu\rho\lambda}(\chi_E)$ are presented as follows:

$$\hat{\alpha}_{\mu\rho}(\alpha) = \alpha\delta_{\mu\rho},$$

$$\hat{\alpha}_{\mu\rho}(\bar{\bar{\alpha}}) = \bar{\bar{\alpha}}(\widehat{W}_\mu \widehat{W}_\rho + \widehat{W}_\rho \widehat{W}_\mu),$$

$$\hat{k}_{\mu\rho\lambda}(\gamma_{E_1}) = i\gamma_{E_1}\delta_{\mu\rho\kappa\lambda}\widehat{W}_\kappa,$$

$$\hat{k}_{\mu\rho\lambda}(\chi_E) = \frac{i\chi_E}{2m}\delta_{\mu\rho\kappa\lambda}\overleftrightarrow{\partial}_\kappa,$$

$$\hat{k}_{\mu\rho\lambda}(\gamma_{E_2}) = -\gamma_{E_2}\widehat{W}_\kappa.$$

In equations is used the definition of covariant spin vector, which is expressed through matrices $\beta_\nu$ according to [22]:

$$\widehat{W}_\mu = -\frac{i}{4m}\delta_{\mu\kappa\delta\eta}\hat{J}^{[\delta\eta]}\overleftrightarrow{\partial}_\kappa,$$

where $\hat{J}^{[\delta\eta]} = \beta_\delta\beta_\eta - \beta_\eta\beta_\delta$. All the derivatives and operators contained in equations, effect on the wave functions $\psi$ and $\bar{\psi}$.

Similarly the tensor (5) is defined, if to enter the constant $\beta, \bar{\bar{\beta}}$, $\gamma_{M_1}$ and $\chi_M$ in (6) – (7), as well as to make the change

$$F_{\nu\mu} \to \tilde{F}_{\nu\mu},$$

where

$$\tilde{F}_{\mu\nu} = \frac{i}{2}\delta_{\mu\nu\rho\sigma}F_{\rho\sigma},$$

and instead of tensor $\hat{k}_{\mu\rho\lambda}(\gamma_{E_2})$ we should enter

$$\hat{k}_{\mu\rho\lambda}(\gamma_{M_2}) = \gamma_{M_2}\widehat{W}_\kappa.$$

We now define the spin structure of the amplitude of Compton scattering on a particle with spin one, taking into account the contributions of the polarizabilities and gyrations on the basis of the Lagrangian (3), following [22]

$$\langle k_2, p_2|\hat{S}|k_1, p_1\rangle = \frac{im\delta(k_1+p_1-k_2-p_2)}{(2\pi)^2\sqrt{4\omega_1\omega_2 E_1 E_2}} M, \tag{9}$$

where M is the amplitude of Compton scattering, which is the sum of the contributions of the polarizabilities and gyrations in accordance with (4) and (5).

As it was shown in [20], the contribution of $\alpha$, $\beta$ and $\bar{\bar{\alpha}}$, $\bar{\bar{\beta}}$ is determined by the sum of the amplitudes

$$M(\alpha,\beta) + M(\bar{\bar{\alpha}},\bar{\bar{\beta}}). \tag{10}$$

Spin structure of $M(\alpha,\beta)$ in (10) has the form:

$$M(\alpha,\beta) = \left(-\frac{2\pi i}{m}\right)\{\alpha\left[F^{(2)}_{\nu\mu}F^{(1)}_{\mu\sigma} + F^{(1)}_{\nu\mu}F^{(2)}_{\mu\sigma}\right] +$$

$$+\beta\left[\tilde{F}^{(2)}_{\nu\mu}\tilde{F}^{(1)}_{\mu\sigma} + \tilde{F}^{(1)}_{\nu\mu}\tilde{F}^{(2)}_{\mu\sigma}\right]\}P_\sigma\bar{\psi}^{(r_2)}(p_2)\beta_\nu\psi^{(r_1)}(p_1). \tag{11}$$

In its turn, the structure of $M(\bar{\bar{\alpha}},\bar{\bar{\beta}})$ is defined in the following way:

$$M(\bar{\bar{\alpha}},\bar{\bar{\beta}}) = \left(-\frac{\pi i}{m}\right)\{\bar{\bar{\alpha}}\left[F^{(2)}_{\nu\mu}F^{(1)}_{\mu\sigma} + F^{(1)}_{\nu\mu}F^{(2)}_{\mu\sigma}\right] +$$

$$+\bar{\bar{\beta}}\left[\tilde{F}^{(2)}_{\nu\mu}\tilde{F}^{(1)}_{\mu\sigma} + \tilde{F}^{(1)}_{\nu\mu}\tilde{F}^{(2)}_{\mu\sigma}\right]\}P_\sigma\bar{\psi}^{(r_2)}(p_2)[\beta_\nu\{\widehat{W}_\mu,\widehat{W}_\rho\} + \{\widehat{W}_\mu,\widehat{W}_\rho\}\beta_\nu]\psi^{(r_1)}(p_1). \tag{12}$$

In equations (11) and (12) we introduce the notations:

$$F^{(2)}_{\nu\mu} = k_{2\nu}e^{(\lambda_2)^*}_\mu - k_{2\mu}e^{(\lambda_2)^*}_\nu,$$

$$F^{(1)}_{\mu\sigma} = k_{1\mu}e^{(\lambda_1)}_\sigma - k_{1\sigma}e^{(\lambda_1)}_\mu,$$

In its turn, $\tilde{F}^{(2)}_{\nu\mu} = \frac{i}{2}\delta_{\nu\mu\varkappa\delta}F^{(2)}_{\varkappa\delta}$, $P_\sigma = \frac{1}{2}(p_1 + p_2)_\sigma$, $p_1$ and $p_2$ are the pulses of the initial and final particles of spin one.

Ten-wave functions in the DKP formalism are represented with the help of the elements of the full matrix algebra $\varepsilon^{AB}$ [25]

$$\psi^{(r)}(p) = \psi^{(r)}_\mu(p)\varepsilon^{\mu 1} + \frac{1}{2}\psi^{(r)}_{[\mu\nu]}(p)\varepsilon^{[\mu\nu]1}.$$

In this ratio

$$\psi^{(r)}_\mu(p) = \frac{i}{\sqrt{2}}\lambda^{(r)}_\mu,$$

$$\psi^{(r)}_{[\mu\nu]}(p) = -\frac{1}{\sqrt{2}m}\left(p_\mu\lambda^{(r)}_\nu - \lambda^{(r)}_\mu p_\nu\right),$$

$\lambda^{(r)}_\mu$ are the components of the polarization vectors of the particle of spin one, and $\varepsilon^{AB}$ are the elements of the full matrix algebra [25]:

$$(\varepsilon^{AB})_{CD} = \delta_{AC}\delta_{BD}, \qquad \varepsilon^{AB}\varepsilon^{CD} = \delta_{BC}\varepsilon^{AD},$$

for a particle of spin 1 indices $A, B, C, D = \mu, [\rho\sigma]$, the square brackets denote antisymmetry on indices $\rho$ and $\sigma$.

Wave functions $\bar{\psi}^{(r)}(p)$ conjugate $\psi^{(r)}(p)$ taking into account the matrix $\eta$ have the form:

$$\bar{\psi}^{(r)}(p) = \psi^+(p)\eta = \left(-\frac{i}{\sqrt{2}}\right)\left[\dot{\lambda}^{(r)}_\mu\varepsilon^{1\mu} + \frac{i}{2m}\varepsilon^{1[\mu\nu]}\left(p_\mu\dot{\lambda}^{(r)}_\nu - p_\nu\dot{\lambda}^{(r)}_\mu\right)\right],$$

where $\dot{\lambda}^{(r)}_\mu\left\{\lambda^{(r)*}_i, \lambda^{(r)}_4\right\}$.

If we use the wave functions $\psi^{(r)}(p)$ and $\bar{\psi}^{(r)}(p)$, we can see that (11) and (12) are consistent with a low-energy representation of Compton scattering amplitude [9,20].

Taking into account the contributions of the spin polarizabilities $L(\gamma_{E_1})$ and $L(\gamma_{M_1})$ to the Lagrangian (3) in the DKP formalism, we get

$$L(\gamma_{E_1}) = -i\frac{\pi}{2m}\gamma_{E_1}\delta_{\mu\rho\varkappa\lambda}F_{\nu\mu}\overleftrightarrow{\partial}_\lambda F_{\rho\sigma}\bar{\psi}(\beta_\nu\widehat{W}_k + \widehat{W}_k\beta_\nu)\overleftrightarrow{\partial}_\sigma\psi, \qquad (13)$$

$$L(\gamma_{M_1}) = -i\frac{\pi}{2m}\gamma_{M_1}\delta_{\mu\rho\varkappa\lambda}\tilde{F}_{\nu\mu}\overleftrightarrow{\partial}_\lambda \tilde{F}_{\rho\sigma}\bar{\psi}(\beta_\nu\widehat{W}_k + \widehat{W}_k\beta_\nu)\overleftrightarrow{\partial}_\sigma\psi. \qquad (14)$$

In the system of the target's rest in approximation when the recoil impulse of a particle is zero, these Lagrangians are defined as follows

$$L(\gamma_{E_1}) = 4\pi\gamma_{E_1}\left(\vec{S}\left[\vec{E}\dot{\vec{E}}\right]\right), \quad (15)$$

$$L(\gamma_{M_1}) = 4\pi\gamma_{M_1}\left(\vec{S}\left[\vec{H}\dot{\vec{H}}\right]\right). \quad (16)$$

We now define the spin structures of the amplitudes based on the contributions of spin polarizabilities $\gamma_{E_1}, \gamma_{M_1}, \gamma_{E_2}, \gamma_{M_2}$ which, according to [6], are bounded with the dipole and quadrupole moments of hadrons respectively. And also we take into account gyrations $\chi_E, \chi_M$:

$$M = M(\gamma_{E_1}, \gamma_{M_1}) + M(\chi_E, \chi_M) + M(\gamma_{E_2}, \gamma_{M_2}).$$

Using the summands of the Lagrangian (4) and (5) $\hat{L}_{\nu\sigma}(\gamma_{E_1})$, $\hat{L}_{\nu\sigma}(\gamma_{E_2})$, $\hat{L}_{\nu\sigma}(\chi_E)$, $\hat{L}_{\nu\sigma}(\gamma_{M_1})$, $\hat{L}_{\nu\sigma}(\gamma_{M_2})$ and $\hat{L}_{\nu\sigma}(\chi_M)$, as well as the previous method for determining the contributions to the amplitude of Compton scattering of the polarizabilities, we get:

$$M(\gamma_{E_1}, \gamma_{M_1}) = i\frac{\pi}{m}(k_1 + k_2)_\lambda \delta_{\mu\rho\lambda\kappa}\{\gamma_{E_1}\left[F_{\nu\mu}^{(2)}F_{\rho\sigma}^{(1)} - F_{\nu\mu}^{(1)}F_{\rho\sigma}^{(2)}\right] +$$

$$+\gamma_{M_1}\left[\tilde{F}_{\nu\mu}^{(2)}\tilde{F}_{\rho\sigma}^{(1)} - \tilde{F}_{\nu\mu}^{(1)}\tilde{F}_{\rho\sigma}^{(2)}\right]\}\bar{\psi}^{(r_2)}(p_2)[\beta_\nu\widehat{W}_k + \widehat{W}_k\beta_\nu]P_\sigma\psi^{(r_1)}(p_1). \quad (17)$$

In the system of the target's rest and in neglecting the recoil of the target's particle, the amplitude (17) takes the form:

$$M(\gamma_{E_1}, \gamma_{M_1}) = -4i\pi(\omega_1 + \omega_2)\omega_1\omega_2\vec{\lambda}^{(r_2)*} \cdot$$

$$\cdot\left\{\gamma_{E_1}\left(\vec{S}[\vec{e}^{(\lambda_2)*}\vec{e}^{(\lambda_1)}]\right) + \gamma_{M_1}\left(\vec{S}\left[[\vec{n}_2\vec{e}^{(\lambda_2)*}][\vec{n}_1\vec{e}^{(\lambda_1)}]\right]\right)\right\}\vec{\lambda}^{(r_1)}. \quad (18)$$

From (17) and (18) it follows that the spin polarizabilities $\gamma_{E_1}$ and $\gamma_{M_1}$ contribute to the amplitude of Compton scattering of a particle of spin one in the third order on the photon energy and, at the same time, the conditions of crossing symmetry and parity conservation relative to space inversion are satisfied.

Within the framework of the above approach, we now define a relativistic invariant lagaranghian and amplitude, which, as it has been shown in [6], consider the contribution of the spin polarizabilities related to electric quadrupole moment of hadrons.

$$L(\gamma_{E_2}) = \frac{\pi\gamma_{E_2}}{2m}\left[(F_{\nu\rho}\overleftarrow{\partial}_k\tilde{F}_{\sigma\rho} + \tilde{F}_{\nu\rho}\overleftarrow{\partial}_\rho F_{\sigma k})\right]\bar{\psi}[\beta_\nu\widehat{W}_k + \widehat{W}_k\beta_\nu]\overleftrightarrow{\partial}_\sigma\psi. \quad (19)$$

Using the Lagrangian (19) S – matrix element we can introduce as follows:

$$S_{fi} = \frac{i\delta(k_1 + p_1 - k_2 - p_2)}{(2\pi)^2 \sqrt{4\omega_1 \omega_2 E_1 E_2}} M.$$

In this ratio the amplitude has the form:

$$M(\gamma_{E_2}) = \frac{\pi \gamma_{E_2}}{2m} P_\sigma \bar{\psi}^{(r_2)}(p_2)[\beta_\nu \widehat{W}_k + \widehat{W}_k \beta_\nu]\psi^{(r_1)}(p_1) \cdot$$

$$\cdot \left[\delta_{\sigma\rho\alpha\beta}\left(k_{2k}F^{(2)}_{\nu\rho}F^{(1)}_{\alpha\beta} - k_{1k}F^{(1)}_{\nu\rho}F^{(2)}_{\alpha\beta}\right) + \delta_{\nu\rho\alpha\beta}\left(k_{2\rho}F^{(2)}_{\sigma k}F^{(1)}_{\alpha\beta} - k_{1\rho}F^{(1)}_{\sigma k}F^{(2)}_{\alpha\beta}\right)\right]. \quad (20)$$

In the system of the target's rest and in neglecting the recoil of the target's particle, the Lagrangian (19) takes the form:

$$L(\gamma_{E_2}) = -4\pi \gamma_{E_2} E_{ik} H_i \hat{S}_k, \quad (21)$$

and the amplitude (20) looks as follows:

$$M(\gamma_{E_2}) = -4\pi \gamma_{E_2} \vec{\lambda}^{(r_2)*} \left[\omega_1 \left(\hat{\vec{S}} \vec{e}^{(\lambda_1)}\right) \left(\vec{e}^{(\lambda_2)*}[\vec{k}_2 \vec{k}_1]\right) + \omega_2 \left(\hat{\vec{S}} \vec{e}^{(\lambda_2)*}\right) \left(\vec{e}^{(\lambda_1)}[\vec{k}_2 \vec{k}_1]\right) - \right.$$

$$\left. -\omega_1 \left(\hat{\vec{S}} \vec{k}_1\right) \left(\vec{k}_2[\vec{e}^{(\lambda_2)*} \vec{e}^{(\lambda_1)}]\right) - \omega_2 \left(\hat{\vec{S}} \vec{k}_2\right) \left(\vec{k}_1[\vec{e}^{(\lambda_2)*} \vec{e}^{(\lambda_1)}]\right)\right] \vec{\lambda}^{(r_1)}. \quad (22)$$

In equations (21) and (22) $\vec{\lambda}^{(r_2)*}$ and $\vec{\lambda}^{(r_1)}$ are the vectors of the polarization of the final and initial particles, and $\vec{e}^{(\lambda_2)*}$ and $\vec{e}^{(\lambda_1)}$ are the similar vectors of the photon polarization, tensor $E_{ik} = \frac{1}{2}(\partial_i E_k + \partial_k E_i)$.

The effective Lagrangian of the two-photon interaction with a particle of spin 1, taking into account the contribution of the spin polarizability associated with the magnetic quadrupole moment of hadrons [6], is defined in this approach as follows:

$$L(\gamma_{M_2}) = -i\frac{\pi \gamma_{M_2}}{2m}\left[(\tilde{F}_{\nu\rho}\overleftarrow{\partial}_k F_{\sigma\rho} + F_{\nu\rho}\overrightarrow{\partial}_\rho \tilde{F}_{\sigma\rho})\right]\bar{\psi}[\beta_\nu \widehat{W}_k + \widehat{W}_k \beta_\nu]\overrightarrow{\partial}_\sigma \psi. \quad (23)$$

Using the Lagrangian (23), we obtain the amplitude of Compton scattering:

$$M(\gamma_{M_2}) = -\frac{\pi \gamma_{M_2}}{m} P_\sigma \bar{\psi}^{(r_2)}(p_2)[\beta_\nu \widehat{W}_k + \widehat{W}_k \beta_\nu]\psi^{(r_1)}(p_1) \cdot$$

$$\cdot \left[\delta_{\nu\rho\alpha\beta}\left(k_{2k}F^{(2)}_{\alpha\beta}F^{(1)}_{\sigma\rho} - k_{1k}F^{(2)}_{\sigma\rho}F^{(1)}_{\alpha\beta}\right) + \delta_{\sigma k\alpha\beta}\left(k_{2\rho}F^{(2)}_{\alpha\beta}F^{(1)}_{\nu\rho} - k_{1\rho}F^{(2)}_{\nu\rho}F^{(1)}_{\alpha\beta}\right)\right]. \quad (24)$$

As follows from (23) and (24), in the system of the target's rest and in neglecting the recoil impulse of the target, effective Lagrangian takes the form:

$$L(\gamma_{M_2}) = 4\pi \gamma_{M_2} H_{ki} \hat{S}_k E_i, \quad (25)$$

and the scattering amplitude is defined as follows:

$$M(\gamma_{M_2}) = -4\pi\gamma_{M_2}\vec{\lambda}^{(r_2)*}\left[\omega_1\left(\hat{\vec{S}}\vec{k}_2\right)\left(\vec{k}_2[\vec{e}^{(\lambda_2)*}\vec{e}^{(\lambda_1)}]\right) + \omega_2\left(\hat{\vec{S}}\vec{k}_1\right)\left(\vec{k}_1[\vec{e}^{(\lambda_2)*}\vec{e}^{(\lambda_1)}]\right) + \right.$$

$$\left. +\omega_1(\vec{k}_2\vec{e}^{(\lambda_1)})\left(\hat{\vec{S}}[\vec{k}_2\vec{e}^{(\lambda_2)*}]\right) - \omega_2(\vec{k}_1\vec{e}^{(\lambda_2)*})\left(\hat{\vec{S}}[\vec{k}_1\vec{e}^{(\lambda_1)}]\right)\right]\vec{\lambda}^{(r_1)}.$$

Tensor $H_{ki}$ in (23) has the form:
$$H_{ki} = \frac{1}{2}(\partial_k H_i + \partial_i H_k).$$

Using the above method of constructing covariant blocks of the effective Lagrangian we obtain the expression for Lagrangian taking into account the electric and magnet giratons respectively

$$L(\chi_E) = -\frac{\pi\chi_E}{4m^2}\delta_{\mu\rho k\lambda}(F_{\nu\mu}\vec{\partial}_\lambda F_{\rho\sigma})\bar{\psi}\beta_\nu\vec{\partial}_\sigma\vec{\partial}_k\psi, \qquad (26)$$

$$L(\chi_M) = -\frac{\pi\chi_M}{4m^2}\delta_{\mu\rho\lambda k}(F_{\nu\mu}\vec{\partial}_\lambda F_{\rho\sigma})\bar{\psi}\beta_\nu\vec{\partial}_k\vec{\partial}_\sigma\psi. \qquad (27)$$

In the non-relativistic approximation for the expressions (26-27) we have

$$L(\chi_E) = 2\pi\chi_E\,(\vec{E}[\vec{\nabla}\vec{E}], \qquad (28)$$

$$L(\chi_M) = 2\pi\chi_M\,(\vec{H}[\vec{\nabla}\vec{H}]. \qquad (29)$$

Considering crossing symmetry and nonconservation of parity, we obtain the second summand of the amplitude, which is determined by the contributions of electric and magnetic gyrations [26]:

$$M(\chi_E,\chi_M) = \frac{2\pi i}{m^2}(k_1+k_2)_\lambda\delta_{\mu\rho\lambda\kappa}\{\chi_E\left[F^{(2)}_{\nu\mu}F^{(1)}_{\rho\sigma} - F^{(1)}_{\nu\mu}F^{(2)}_{\rho\sigma}\right] +$$

$$+\chi_M\left[\tilde{F}^{(2)}_{\nu\mu}\tilde{F}^{(1)}_{\rho\sigma} - \tilde{F}^{(1)}_{\nu\mu}\tilde{F}^{(2)}_{\rho\sigma}\right]\}P_k P_\sigma\bar{\psi}^{(r_2)}(p_2)\beta_\nu\psi^{(r_1)}(p_1). \qquad (30)$$

If in the equation (30) we use the approximation $\vec{P} = 0$, i.e. when the particle is at rest and we neglect its recoil impulse, then from (30) in this case it follows:

$$M(\chi_E,\chi_M) = 4\pi\omega_1\omega_2(\vec{\lambda}^{(r_2)*}\vec{\lambda}^{(r_1)})\{\chi_E(\vec{k}_1+\vec{k}_2)[\vec{e}^{(\lambda_2)*}\vec{e}^{(\lambda_1)}] +$$

$$+\chi_M(\vec{k}_1+\vec{k}_2)[\vec{\Sigma}_2\vec{\Sigma}_1]\},$$

where $\vec{\Sigma}_2 = [\vec{n}_2 \vec{e}^{(\lambda_2)*}]$, $\vec{\Sigma}_1 = [\vec{n}_1 \vec{e}^{(\lambda_1)}]$.

### Tensor spin polarizabilities of the spin 1 particles

We now define the relativistically invariant Lagrangian blocks that are associated with the spin tensor polarizabilities of the particle of spin 1.

We construct at first the part of the Lagrangian, which is determined by the electromagnetic field tensor $F_{\nu\mu}$ and the covariant spin vector $\widehat{W}_\mu$

$$\widehat{L}(\bar{\bar{\alpha}}_E) = \pi \bar{\bar{\alpha}}_E (F_{\nu\rho} \overleftrightarrow{\partial}_\lambda F_{\lambda\sigma} + F_{\nu\lambda} \overrightarrow{\partial}_\lambda F_{\rho\sigma}) \bar{\psi}[\beta_\nu \{\widehat{W}_\mu, \widehat{W}_\rho\} + \{\widehat{W}_\mu, \widehat{W}_\rho\}\beta_\nu]\psi. \quad (31)$$

In the particle rest frame, and neglecting the the recoil impulse, we can use the transition

$$\bar{\psi}[\beta_\nu \{\widehat{W}_\mu, \widehat{W}_\rho\} + \{\widehat{W}_\mu, \widehat{W}_\rho\}\beta_\nu]\psi \to$$

$$\to (-2)\left[\left(\vec{\lambda}_a^{(r)*}\vec{\lambda}_d^r + \vec{\lambda}_d^{(r)*}\vec{\lambda}_a^r\right) - \delta_{ad}(\vec{\lambda}^{(r)*}\vec{\lambda}^{(r)})\right] = (-2)[\delta_{ad} - \{\hat{S}_a, \hat{S}_d\}], \quad (32)$$

where the indices a, d take the values 1,2,3.

Using expression (32) and the tensor $F_{\nu\mu}$ components we obtain

$$L(\bar{\bar{\alpha}}_E) = -2i\pi\bar{\bar{\alpha}}_E \left[(\dot{\vec{E}}\vec{E}) + (\vec{H}\dot{\vec{H}}) - \{\dot{\vec{E}}\hat{\vec{S}}, \vec{E}\hat{\vec{S}}\} + \delta_{ikl}(\partial_i E_j)H_l\{\hat{S}_j, \hat{S}_k\}\right]. \quad (33)$$

In turn, the part of the Lagrangian, which is determined by dual electromagnetic tensors, has the form

$$\widehat{L}(\bar{\bar{\alpha}}_M) = \pi \bar{\bar{\alpha}}_M (\tilde{F}_{\nu\rho} \overleftrightarrow{\partial}_\lambda \tilde{F}_{\lambda\sigma} + \tilde{F}_{\nu\lambda} \overrightarrow{\partial}_\lambda \tilde{F}_{\rho\sigma}) \bar{\psi}[\beta_\nu \{\widehat{W}_\rho, \widehat{W}_\sigma\} + \{\widehat{W}_\rho, \widehat{W}_\sigma\}\beta_\nu]\psi. \quad (34)$$

In the nonrelativistic approximation of expression (34) it follows:

$$L(\bar{\bar{\alpha}}_M) = -2i\pi\bar{\bar{\alpha}}_M \left[(\vec{H}\dot{\vec{H}}) + (\dot{\vec{E}}\vec{E}) - \{\dot{\vec{H}}\hat{\vec{S}}, \vec{H}\hat{\vec{S}}\} + \delta_{ikl}(\partial_i H_j)E_l\{\hat{S}_j, \hat{S}_k\}\right]. \quad (35)$$

The Compton scattering amplitude is obtained on the basis of the Lagrangian (31) and is defined as follows:

$$M(\bar{\bar{\alpha}}_E) = i\pi\bar{\bar{\alpha}}_E \bar{\psi}^{(r_2)}(p_2)[\beta_\nu \{\widehat{W}_\rho, \widehat{W}_\sigma\} + \{\widehat{W}_\rho, \widehat{W}_\sigma\}\beta_\nu]\psi^{(r_1)}(p_1) \cdot$$

$$\cdot \left[\left(-k_{2\lambda}F^{(2)}_{\nu\rho}F^{(1)}_{\lambda\sigma} + k_{1\lambda}F^{(1)}_{\nu\rho}F^{(2)}_{\lambda\sigma}\right) + \left(-k_{2\lambda}F^{(1)}_{\nu\lambda}F^{(2)}_{\rho\sigma} + k_{1\lambda}F^{(2)}_{\nu\lambda}F^{(1)}_{\rho\sigma}\right)\right]. \quad (36)$$

To get a piece of the Compton scattering amplitude, which follows from the Lagrangian (34), it is enough (36) to change $F_{\nu\mu}^{(2)}$ for $\tilde{F}_{\nu\mu}^{(2)}$, and to replace $F_{\mu\nu}^{(1)}$ by $\tilde{F}_{\mu\nu}^{(1)}$

If we confine ourselves in (36) to the third order in the photon energy, the amplitude (36) takes the form:

$$M(\bar{\bar{\alpha}}_E) = 2\pi\bar{\bar{\alpha}}_E\omega^3\left[(\hat{k}_2\vec{e}^{(\lambda_1)})\left(\hat{\vec{S}}\vec{e}^{(\lambda_2)*}\cdot\hat{\vec{S}}\hat{k}_1 + \hat{\vec{S}}\hat{k}_1\cdot\hat{\vec{S}}\vec{e}^{(\lambda_2)*}\right) - \right.$$

$$\left. - (\hat{k}_1\vec{e}^{(\lambda_2)*})\left(\hat{\vec{S}}\vec{e}^{(\lambda_1)}\cdot\hat{\vec{S}}\hat{k}_2 + \hat{\vec{S}}\hat{k}_2\cdot\hat{\vec{S}}\vec{e}^{(\lambda_1)}\right)\right],$$

where $\hat{k}_1 = \frac{\vec{k}_1}{|\vec{k}_1|}$, $\hat{k}_2 = \frac{\vec{k}_2}{|\vec{k}_2|}$.

Lagrangians associated with parity violation can be determined using the expressions (31) and (34). Since the Lagrangian associated with the pseudoscalarity $\widehat{W}_\mu$, can be represented as follows:

$$L(\bar{\bar{\chi}}_E) = \frac{\pi\bar{\bar{\chi}}_E}{2m}\left[(F_{\nu\rho}\overleftarrow{\partial}_\lambda F_{\lambda\sigma} + F_{\nu\lambda}\overrightarrow{\partial}_\lambda F_{\rho\sigma})\right]\bar{\psi}[\beta_\nu\widehat{W}_\rho + \widehat{W}_\rho\beta_\nu]\overleftrightarrow{\partial}_\sigma\psi. \quad (37)$$

In the non-relativistic approximation, it has the form:
$$L(\bar{\bar{\chi}}_E) = 4\pi\bar{\bar{\chi}}_E(\vec{E}\vec{\partial})(\vec{E}\hat{\vec{S}}). \quad (38)$$

Similarly, the Lagrangian on the basis of (34) is determined

$$L(\bar{\bar{\chi}}_M) = \frac{\pi\bar{\bar{\chi}}_M}{2m}\left[(\tilde{F}_{\nu\rho}\overleftarrow{\partial}_\lambda\tilde{F}_{\lambda\sigma} + \tilde{F}_{\nu\lambda}\overrightarrow{\partial}_\lambda\tilde{F}_{\rho\sigma})\right]\bar{\psi}[\beta_\nu\widehat{W}_\rho + \widehat{W}_\rho\beta_\nu]\overleftrightarrow{\partial}_\sigma\psi, \quad (39)$$

which in a non-relativistic limit can be represented as follows:

$$L(\bar{\bar{\chi}}_M) = 4\pi\bar{\bar{\chi}}_M(\vec{H}\vec{\partial})(\vec{H}\hat{\vec{S}}). \quad (40)$$

### Conclusion

In this paper is proposed the method of relativistic-invariant effective tensor representation of decomposition Lagrangian and the amplitudes of the two-photon interaction with hadrons on the photon energy.

Within the framework of the formalism of the Duffin-Kemmer-Petiau on the basis of the method the authors have set the spin polarizabilities of the spin 1 particles, which are characteristic for the hadron spin 1/2, as well as they have obtained the new spin polarizabilities for particles of spin 1 and associated with the presence of the tensor polarizabilities of these particles.

It is shown that in the proposed covariant approach, taking into account cross-symmetry, the laws of conservation of parity and the gauge invariance, certain spin polarizabilities and gyrations of the particle of spin one contribute to the expansion of the the Compton scattering amplitude, beginning with the relevant orders of the photon energy in accordance with the low-energy theorems for the process.

[1] Raguza, S. Third-order spin polarizabilities of the nucleon: I / S. Raguza // Phys. Rev. D. – 1993. – Vol. 47. - № 9. – P. 3757 – 3767.

[2] Raguza, S. Third-order spin polarizabilities of the nucleon: II / S. Raguza // Phys. Rev. D. – 1994. – Vol. 49. - № 7. – P. 3157 – 3159.

[3] Hill, R.J. The NRQED lagrangian at order $\frac{1}{M^4}$/ R.J. Hill, G. Lee, G. Paz, M.P. Solon // Phys. Rev. D. – 2013. – Vol. 87. –№5. - P. 053017-1-13.

[4] B. R. Holstein and S. Scherer, Hadron Polarizabilities [Electronic resource]. – 2013. – Mode of access: http: // hep-ph/1401.0140v1. – Date of access: 31.12.2013].

[5] C. E. Carlson and M. Vanderhaeghen, Constraining off-shell effects using low-energy Compton scattering [Electronic resource]. – 2011. – Mode of access: http://physics.atom-ph/1109.3779. – Date of access: 04.10.2011.

[6] Low-energy Compton scattering of polarized photons on polarized nucleons / D. Babusci [et. al.] // Rev. C. – 1998. – Vol. 58. – P. 1013 – 1041.

[7] Chen, J. W. The polarizability of the deuteron / J. W. Chen [et. al.] // Nucl. Phys. A. – 1998. – Vol. 644. – P. 221 – 234.

[8] Friar, J. L. Deuteron dipole polarizability and sum rules / J. L. Friar, G. L. Payne // Rev. C. – 2005. – Vol. 72. – P. 014004-1 –014004-6.

[9] Lin, K.Y. Forward dispersion relation and low-energy theorems for Compton scattering on spin – 1 targets / K.Y. Lin, J.C. Chen // J. Phys. G: Nucl. Phys. – 1975 – Vol.1 – P. 394 – 399.

[10] F. Hagelstein. Sum Rules for Electromagnetic Moments and Polarizabilities of Spin-1 Particles in Massive Yang-Mills QED / Masterarbeit in Physikvorgelegt dem Fachbereich Physik, Mathematik und Informatik (FB 08) der Johannes Gutenberg-Universit.at Mainz am 11. März 2014.

[11] Bedaque, P. F. Parityviolationin $\gamma p$ Compton Scattering / P. F. Bedaque, M. J. Savage // Phys. Rev. C. – 2000 – V. 62 – P. 018501 – 1 – 6.

[12] Gorchtein, M. Forward Compton Scattering with neutral current: constraints from sum rules / M. Gorchtein, X. Zhang [Electronic resourse]. – 2015. Mode of access: http: // nucl-th / 1501.0535v1. – Date of access: 22.01.2015.

[13] F. I. Fedorov, Theory of Gyrotropy [in Russian], Nauka i Tekhnika, Minsk (1976).

[14] Ilyichev, A. Static polarizability vertex and itsapplications / A. Ilyichev, S. Lukashevich, N. Maksimenko // [Electronicresource]. – 2006. – Modeofaccess:arXiv: // hep-ph/0611327v1. – Dateofaccess: 27.11.2006.

[15] Zhang, Y. Proton Compton scattering in a unified proton - $\Delta^+$Model / Y. Zhang, K. Savvidy // Phys. Rev. C. – 2013. – Vol. 88. –P. 064614– 1 – 12.

[16] Krupina, N. Separation of proton polarizabilities with the beam asymmetry of Compton scattering / N.Krupina, V. Pascalutsa // Phys. Rev. Lett. – 2013. – Vol.110. – №26. - P. 262001 – 1 – 4.
[17] N. V. Maksimenko, Dokl. Akad. Nauk Belarusi, **36**, No. 6, 508–510 (1992).
[18] V. V. Andreev, O. M. Deryuzhkova, and N. V. Maksimenko, Problemy Fiziki, Matematiki i Tekhniki [Problems of Physics, Mathematics, and Physics], No. 3(20), 7–12 (2014).
[19] E. V. Vakulina and N. V. Maksimenko, Problemy Fiziki, Matematiki i Tekhniki [Problems of Physics, Mathematics, and Physics], No. 3, 16–18 (2013).
[20] Maksimenko, N.V. Spin1 Particle Polarizability in the Duffin–Kemmer–Petiau Formalism / N.V. Maksimenko, E.V. Vakulina, S.M. Kuchin // Physics of Particles and Nuclei Letters. – 2015. – Vol. 12, №. 7, P. 807–812
[21] Maksimenko, N.V. Dipole spin polarizabilities and gyrations of spin-one particles in the Duffin–Kemmer–Petiau formalism / N.V. Maksimenko, E.V. Vakulina, S.M. Kuchin // Higher Education. physics. – 2016. – Vol. 59. - № 6.
[22] A. A. Bogush, Introduction to Gauge Field Theory of Electroweak Interactions [in Russian], Nauka i Tekhnika, Minsk (1987).